\RequirePackage[hyphens]{url}

\documentclass[11pt]{article}
\usepackage[a4paper]{geometry}
\usepackage[numbers]{natbib}

\usepackage[utf8]{inputenc} 
\usepackage{helvet}

\renewcommand{\baselinestretch}{1.2} 
\usepackage{hyperref}       
\usepackage{url}            
\usepackage{booktabs}       
\usepackage{amsfonts}       
\usepackage{nicefrac}       
\usepackage{microtype}      
\usepackage{authblk}
\usepackage{graphicx}
\usepackage{amsmath}
\usepackage{subfig}
\usepackage{float}
\title{Exeum: A Decentralized Financial Platform for Price-Stable Cryptocurrencies}

\author{Jaehyung Lee}
\author{Minhyung Cho}
\affil{\{jlee, minhyung.cho\}@exeum.org}

\affil{Exeum Foundation\\ \url{https://www.exeum.org}}
\date{}
\begin{document}

\maketitle

\begin{abstract}
Price stability has often been cited as a key reason that cryptocurrencies have not gained widespread adoption as a medium of exchange and continue to prove incapable of powering the economy of decentralized applications (DApps) efficiently. Exeum proposes a novel method to provide price stable digital tokens whose values are pegged to real world assets, serving as a bridge between the real world and the decentralized economy.

Pegged tokens issued by Exeum -- for example, USDE refers to a stable token issued by the system whose value is pegged to USD -- are backed by virtual assets in a virtual asset exchange where users can deposit the base token of the system and take long or short positions. Guaranteeing the stability of the pegged tokens boils down to the problem of maintaining the peg of the virtual assets to real world assets, and the main mechanism used by Exeum is controlling the swap rate of assets. If the swap rate is fully controlled by the system, arbitrageurs can be incentivized enough to restore a broken peg; Exeum distributes statistical arbitrage trading software to decentralize this type of market making activity. The last major component of the system is a central bank equivalent that determines the long term interest rate of the base token, pays interest on the deposit by inflating the supply if necessary, and removes the need for stability fees on pegged tokens, improving their usability.

To the best of our knowledge, Exeum is the first to propose a truly decentralized method for developing a stablecoin that enables 1:1 value conversion between the base token and pegged assets, completely removing the mismatch between supply and demand. In this paper, we will also discuss its applications, such as improving staking based DApp token models, price stable gas fees, pegging to an index of DApp tokens, and performing cross-chain asset transfer of legacy crypto assets.

\end{abstract}

\newpage

\renewcommand{\baselinestretch}{1.15}
\tableofcontents
\renewcommand{\baselinestretch}{1.2}

\newpage

\section{Introduction}
\label{sec:intro}

Bitcoin \cite{nakamoto2008bitcoin}, the first digital asset with no centralized issuer or controller, has demonstrated that it is possible to practically solve a distributed consensus problem under the presence of Byzantine failure by implementing a built-in incentive structure that makes attacks economically implausible due to cost. The implications have been profound, eventually leading us to the era of smart contract and decentralized application (DApp) platforms (e.g. Ethereum \cite{ethereum}) by extending the concepts introduced in Bitcoin further and recording the state variables of generic program execution on a distributed ledger. 

However, the original purpose of Bitcoin, becoming the first decentralized digital currency powering a decentralized economy, still remains elusive. The major hurdle is that Bitcoin does not have a stable price and fails to become a unit of account, which is crucial for the efficiency of market economics based on it. Using currency as a unit of account essentially requires social agreement, and price stability is more a consequence than a cause. Even in the best case, whereby Bitcoin obtains this social agreement, it cannot properly serve the decentralized economy since it shares the same deficit as the gold standard system -- the total amount of Bitcoin is fixed. As the total amount of world economic activity increases, the use of Bitcoin  will eventually result in a recession as evidenced by the Great Depression occurring as a consequence of the gold standard system.

To support a stable decentralized economy, the design of a digital currency should have a fiat-like stabilization mechanism. The supply of such a currency needs to be carefully adjusted to keep the balance between supply and demand, and to maintain the stability of the economy built upon it. One way of achieving this is to create a decentralized equivalent of the Federal Reserve System (Fed). The system could algorithmically control the supply of the currency in various ways and intervene in the market to stabilize the economy. However, this is only possible after the decentralized economy has fully materialized, with automatically measurable indicators. Moreover, the task itself may be too dynamic and complicated to be handled with a prescribed algorithm -- we eventually need a system with strong learning capabilities.

A more realistic, step-by-step approach to this problem is to delegate the task of monetary policy to trusted centralized organizations (akin to the Fed), and devise a decentralized mechanism to peg a digital token to a fiat currency. Several independent approaches have been proposed to achieve this; they can be categorized as IOU-backed (e.g.Tether \cite{tether} and TrueUSD \cite{trueusd}), crypto-backed (e.g. Bitshares \cite{bitshares} and MakerDAO \cite{dai}), and Seigniorage share \cite{seign} based (e.g. Basis \cite{basis} and Carbon \cite{carbon}) models. However, we offer a detailed analysis in Section~\ref{sec:comparison} of why there has not been a satisfactory design for stable fiat-pegged and fully decentralized digital tokens.

Our approach to stablecoins is based on financial derivatives in the real world, which was briefly examined in the Ethereum white paper \cite{ethereum}. In fact, the first two of the three approaches cited above can be interpreted as variations of financial derivatives. IOU-backed models work exactly in the same way gold warrants do -- just like gold warrants are issued by mining companies or mints, a crypto exchange (Tether) or trust companies (TrueUSD) issue stablecoins backed by fiat deposit. Users can own fiat pegged tokens instead of fiat money, which parallels owning gold warrants instead of physical gold.

However, this approach is neither decentralized nor scalable. What happens if more than half of total issued USD is used as collateral and stored in those trust companies? It will inevitably lead to regulatory constraints and the supply of those derivatives will be limited to a certain extent, which is clearly problematic for the decentralized economy; if a DApp requires users to pay using TrueUSD and there is no way to obtain it, then the DApp will have no users. 

\begin{figure}
  \centering
    \includegraphics[width=\textwidth]{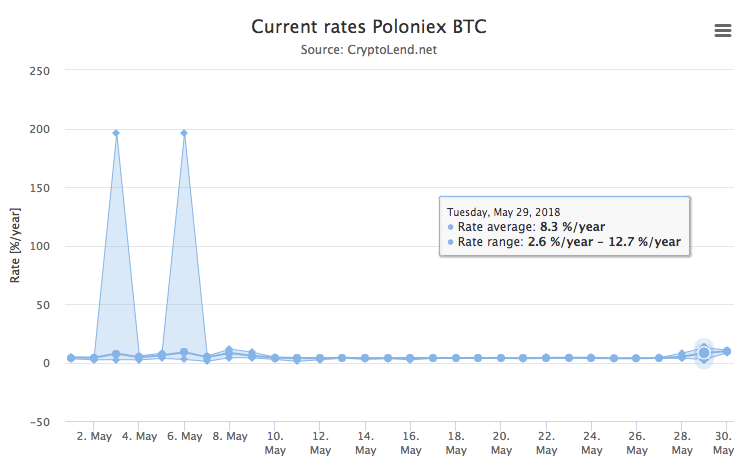}
  \caption{Bitcoin lending rate history on Poloniex}
  \label{fig:lending}
\end{figure}

Crypto-backed models avoid centralization and regulation risk by utilizing smart contracts -- thousands of individuals deposit collateral in cryptocurrencies and issue fiat derivatives in a collective way. However, it has one problem which is not easy to overcome -- collateral is prone to a significant amount of market risk, and individuals need to get paid for exposure to that risk. Considering that fiat derivatives need to be overcollateralized by at least 2 times to avoid liquidation and collateral needs to be locked for safety, derivative holders need to pay a significant amount of `stability' fees to issuers. Lending Bitcoin on margin-enabled exchanges yields 2-10\% profit per year (see Figure~\ref{fig:lending}). Setting aside the 1x liquidity obtained by the issued fiat derivative, a proper fee rate for overcollateralized fiat derivatives would still be up to 2-10\% per year, which is prohibitive for daily usage as fiat pegged tokens.

We focus on this efficiency problem and try to reduce the burden of stablecoin issuers as much as possible by implementing a decentralized derivatives exchange with margin trading enabled. Users can detach their crypto deposit from individual contracts and utilize margin trading to make the best use of their funds. Derivatives in this exchange can be best understood as virtual assets. A trade pair can be regarded as a virtual asset only if the quote asset is accepted as deposit by the exchange. For instance, if the exchange accepts BTC as deposit, a long position in USD/BTC pair can represent a virtual USD (base asset) quoted in BTC (quote asset). Users can margin trade virtual assets by taking long or short positions within their total deposit. As we will see in Section~\ref{sec:currency_service}, if the deposit is made in the quote asset, a fully backed long position is free from the risk of margin calls, making it possible to issue a new value pegged token backed by that position.

In the rest of this paper, we describe the architecture of Exeum, explain the functions of key components, provide a stability analysis to understand the robustness of the system in various scenarios, and analyze other approaches in relation to the proposed design. A number of interesting applications of Exeum service are proposed, and to conclude the paper, we discuss various implications of this work on the future of decentralized and traditional economies.

\section{Exeum}
\label{sec:exeum}

Exeum is a service implemented on top of a DApp platform that provides a stable digital token in a fully decentralized and robust manner. First, we create a decentralized virtual asset exchange where virtual assets whose values are pegged to the values of real world assets (e.g. gold, USD, or Bitcoin) are traded (long or short positions may be taken with Exeum token, the base token of the service, as deposit). Second, we implement a service to issue value stable tokens with various properties, using the virtual assets as collateral. Once properly tokenized, end users may remain indifferent to underlying mechanics while holding and using these stable tokens.

\subsection{The Design}

\begin{figure}
  \centering
    \includegraphics[width=\textwidth]{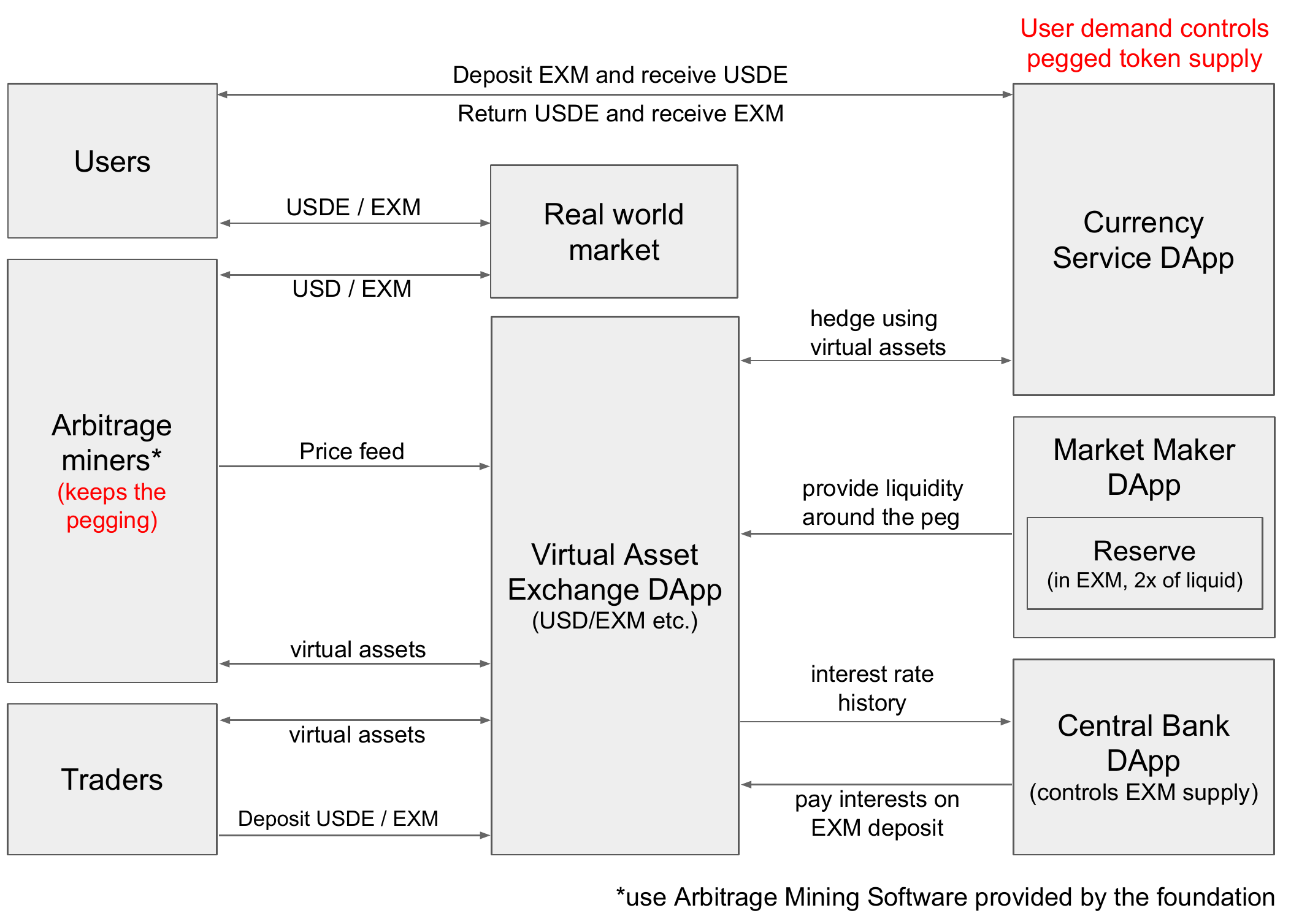}
  \caption{The architecture of Exeum}
  \label{fig:exeum}
\end{figure}

In this section, we present the design of the Exeum service, which is meant to be implemented on top of a DApp platform with its own base token (named Exeum token, EXM). Figure~\ref{fig:exeum} shows the overall architecture of the service, which is composed of five major components: Virtual Asset Exchange DApp, Central Bank DApp, Currency Service DApp, Market Maker DApp, and Arbitrage Mining Software.

In the following sections, each component of the system is described in detail and user flows are illustrated with examples.

\subsection{Virtual Asset Exchange DApp}

A virtual asset is defined as a trade pair whose quote asset is accepted as deposit by a derivatives exchange without an expiry date (i.e. Contract For Difference in financial terms). Virtual assets have the same notional value as their underlying assets. The Virtual Asset Exchange DApp implements a virtual asset exchange in a fully decentralized way to eliminate a significant amount of counterparty risk. From the user’s perspective, it is not much different from centralized exchanges providing margin trading and crypto derivatives (e.g. BitMEX \cite{bitmex}). Users can log in using the web interface, deposit Exeum, see order books, and place long / short orders. The exchange accepts Exeum tokens (EXM) as a margin deposit (i.e. EXM is used as a quote asset) for major virtual assets utilized for backing up pegged tokens. The main reason for doing so is to remove the necessity of requiring stability fees on pegged tokens thanks to the monetary policy of the Central Bank DApp, as to be explained in Section~\ref{sec:central_bank}. Another reason is that more liquidity can be provided for those assets by Market Maker DApp, which utilizes the initial EXM reserve for market making. It is also possible to open a market using pegged tokens (e.g. USDE) as quote assets since pegged tokens are fungible and general-purpose. For many users, it is still more intuitive to deposit and settle in USDE than Bitcoin or Exeum tokens, however, issuing pegged tokens using the virtual assets on this market is potentially risky and needs further analysis.

\begin{figure}
  \centering
    \includegraphics[width=\textwidth]{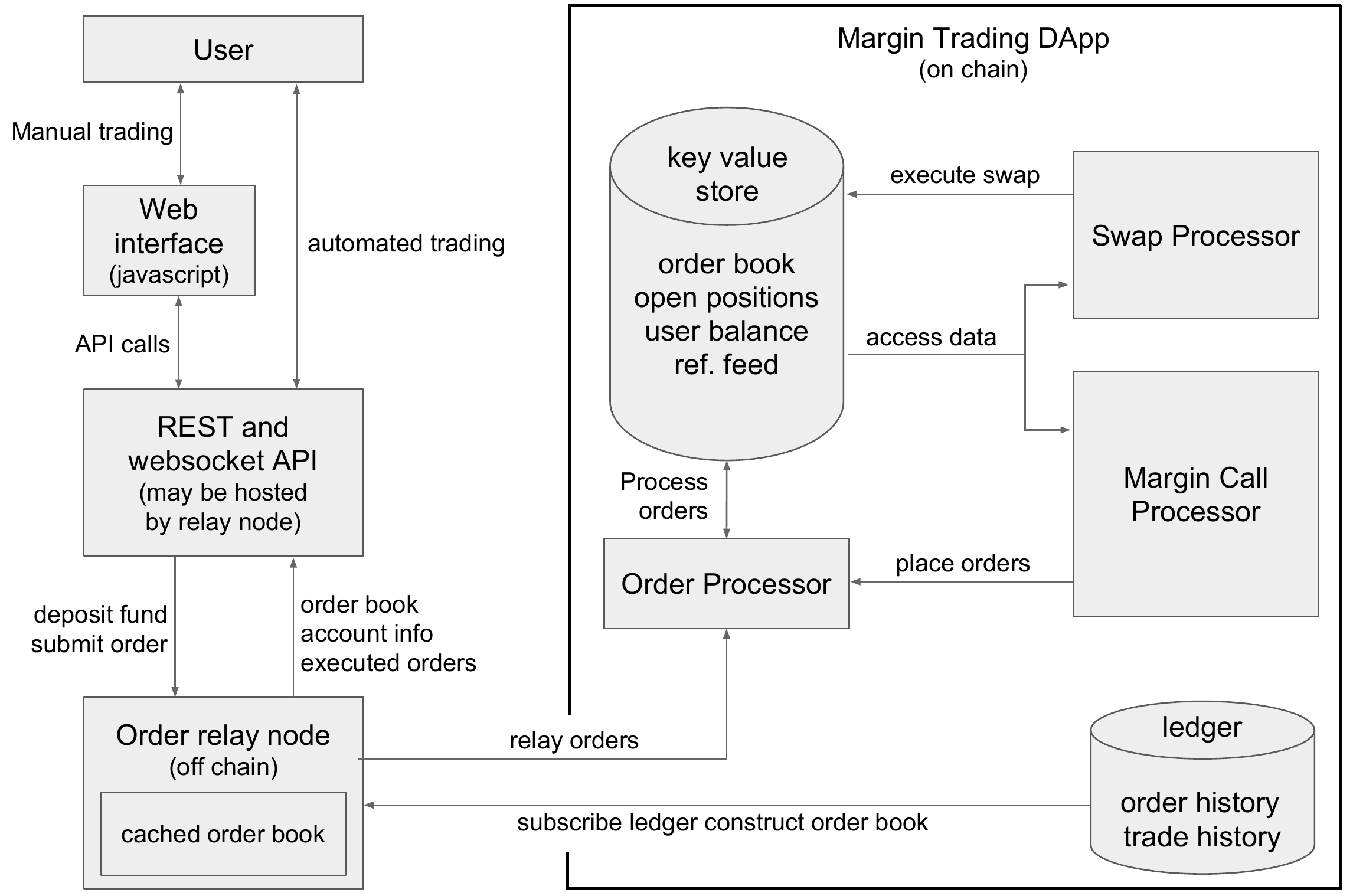}
  \caption{Architecture of Virtual Asset Exchange DApp}
  \label{fig:exchange_dapp}
\end{figure}

All user deposits are stored on the DApp smart contract account and only accessible via smart contracts, removing the necessity of a trusted third party. Implementing a completely decentralized exchange backend using currently available public DApp platforms, such as Ethereum, will require a significant amount of engineering effort due to various well-publicized limitations and we anticipate the need to eventually implement an off-chain solution to satisfy the scalability and security requirements. Figure~\ref{fig:exchange_dapp} illustrates the architecture of the Virtual Asset Exchange DApp. In the proposed architecture, DApp platforms are assumed to have an improved transaction rate as compared to current throughput capabilities so that at least order books can be maintained on chain -- otherwise, the system cannot avoid a remarkable delay between order submission and execution, degrading user experience drastically. Upcoming platform improvements such as Ethereum with Casper / sharding / Plasma and ZILLIQA \cite{zilliqa} may satisfy this requirement.

\subsubsection{Margin Trading DApp}
\label{sec:margin_trading_dapp}
The Margin Trading DApp, implemented as an on-chain DApp, is composed of a key value store and a number of processors -- Order Processor, Swap Processor, and Margin Call Processor. All the processors can access the key value store and the blockchain ledger.

The Order Processor stores order books and positions of each trader on the key value store efficiently, approves order submissions, and executes orders in a batch manner. Order execution is immediate and does not need a separate settlement process since assets being traded are virtual -- changes in user positions and balances are reflected immediately upon order execution. The Order Processor also processes trade fees -- for each executed order, the trader who has taken the order pays a taker fee to the system, then the system pays a rebate to the trader who has placed the order. The rebate amount is slightly smaller than the amount of the taker fee; the difference is taken by the system and used for system operations. The rebate encourages market making activity, thus providing more liquidity to market participants.

How order execution is implemented affects the fairness of the system significantly. Orders submitted by traders will compete against each other at the milliseconds level, and there is a chance of centralization or oligarchy if only the market makers with the fastest order submissions survive. To avoid that scenario, our implementation can execute orders in batches, and partially execute all the competing orders in the same batch proportional to the order amount. In this way, the opportunity can be shared and relatively slower players may survive. 

The Margin Call Processor periodically checks conditions for margin calls, notifies traders, and places margin call orders when conditions are met. If the DApp platform does not support scheduled transactions (for example, Ethereum does not), updates from external feeds can be used to trigger an additional transaction to execute the Margin Call Processor. To prevent attackers from manipulating the market price for brief periods and profiting from that manipulation, the Margin Call Processor needs to monitor external market prices. To safely obtain such a feed, we propose a method: obtaining the market price feed from the Arbitrage Mining Software. Since the software requires users to deposit base tokens on the virtual asset exchange, a proper design can make it cost effective and resistant to Sybil attacks (refer to Section~\ref{sec:arbitrage} for details).

The maximum amount of leverage provided by the exchange should be set within an appropriate range to maximize market maker profit and maintain system stability at the same time. One way is to dynamically adjust the maximum leverage to keep the probability of catastrophic margin calls low. This topic will be revisited in Section~\ref{sec:stability} and Section~\ref{sec:discussion}.

The Swap Processor analyzes the trade history, determines the proper interest rate for each of the traded virtual assets for maintaining the peg, and executes swaps on all the open positions periodically. Assigning higher interest rates for less preferred assets by traders incentivizes the long position of those assets and the Swap Processor may adjust borrowing and lending rates of assets to keep the whole system profitable.

\subsubsection{Off-Chain Relay Node}

With the recent advancements in blockchain technology, it is highly likely that leading DApp platforms such as Ethereum and ZILLIQA can support tens of thousands of transactions per second in the relatively near future -- which would be enough for maintaining on-chain order books for a number of currency pairs. However, it is not clear how to motivate operator nodes to serve read only requests, and more importantly, data streams based on permanent connections. For this reason, our design provides off-chain relay nodes for serving cached order books and exchange API calls, with a reward mechanism so that relay nodes can function as independent asset exchange websites with localized user interfaces. If Exeum is implemented using a layer 2 design like Plasma, relay nodes may function as block producers on the Plasma chain and the design below may be simplified or modified accordingly.

Exeum relay nodes need to monitor transactions on the DApp platform and maintain up-to-date order books and user positions. They can also validate order submission and forward to the platform. In doing so, relay nodes take a small fee per forwarded order to remain profitable. To prevent abuse, relay nodes need to support a sufficient number of users prior to receiving relay fees. The Margin Trading DApp will periodically update the list of relay nodes meeting minimum thresholds for traffic and distribute fees accordingly. In case a relay node has failed to reach a sufficient traffic level, the DApp retains and accumulates the fee until the relay reaches the traffic threshold. If a user does not specify a relay and sends the transaction on her own, the DApp takes the fee and sends to its own reserve for market making (please refer to Section~\ref{sec:market_maker}). In this way, users are not incentivized to skip relay nodes and submit transactions on their own.

A potential issue is that relay nodes become compromised and engage in front running or censor incoming orders. However, since the same on-chain order book is shared across relay nodes, users can simply boycott misbehaving relay nodes and use honest ones. Note that in case of Ethereum, existing DEX protocols such as 0x \cite{0x} assume an off-chain order book is maintained by relay nodes and there is no way for users to switch to other relays to access the liquidity.

\subsubsection{Adjusting Swap Rates}
\label{sec:swap_rates}

In the virtual asset exchange, a trader taking a long position of USD/EXM within her margin amount can be interpreted as borrowing EXM from the system, purchasing USD, and lending the purchased USD back to the system. The system needs to pay the lending interest of USD to the trader and collect the borrowing interest of EXM from her.

The borrowing/lending interest rates of all the assets are determined by the Swap Processor to maintain the peg and keep the Exeum service functional. It computes the volume weighted difference of the index price and the price of executed orders over a predetermined period (e.g. 4 hours). If the price of executed orders is greater than the index price, the lending rate of the base asset needs to decrease, or the borrowing rate of the quote asset needs to increase. The processor optimizes the borrowing and lending rate of assets over all the trading pairs available. The borrowing and lending rate of an asset can be the same, in which case the swap is made on a peer-to-peer basis. However, if necessary, the system may lower the lending rate slightly and collect the difference to fund its own operation.

\subsection{Arbitrage Mining Software}
\label{sec:arbitrage}

Arbitrage Mining Software is freely distributed market making software developed and maintained by Exeum and run by arbitrage miners, a decentralized group of market makers protecting the peg of the virtual asset exchange. Incentivizing market making by fair policies may not be enough for decentralization -- there is a risk that the system is controlled by a group of centralized market makers similar to how Bitcoin mining has become centralized. Our design tries to maximize the participation of the general public by providing the right tool and incentive structure.

\begin{figure}
  \centering
    \includegraphics[width=\textwidth]{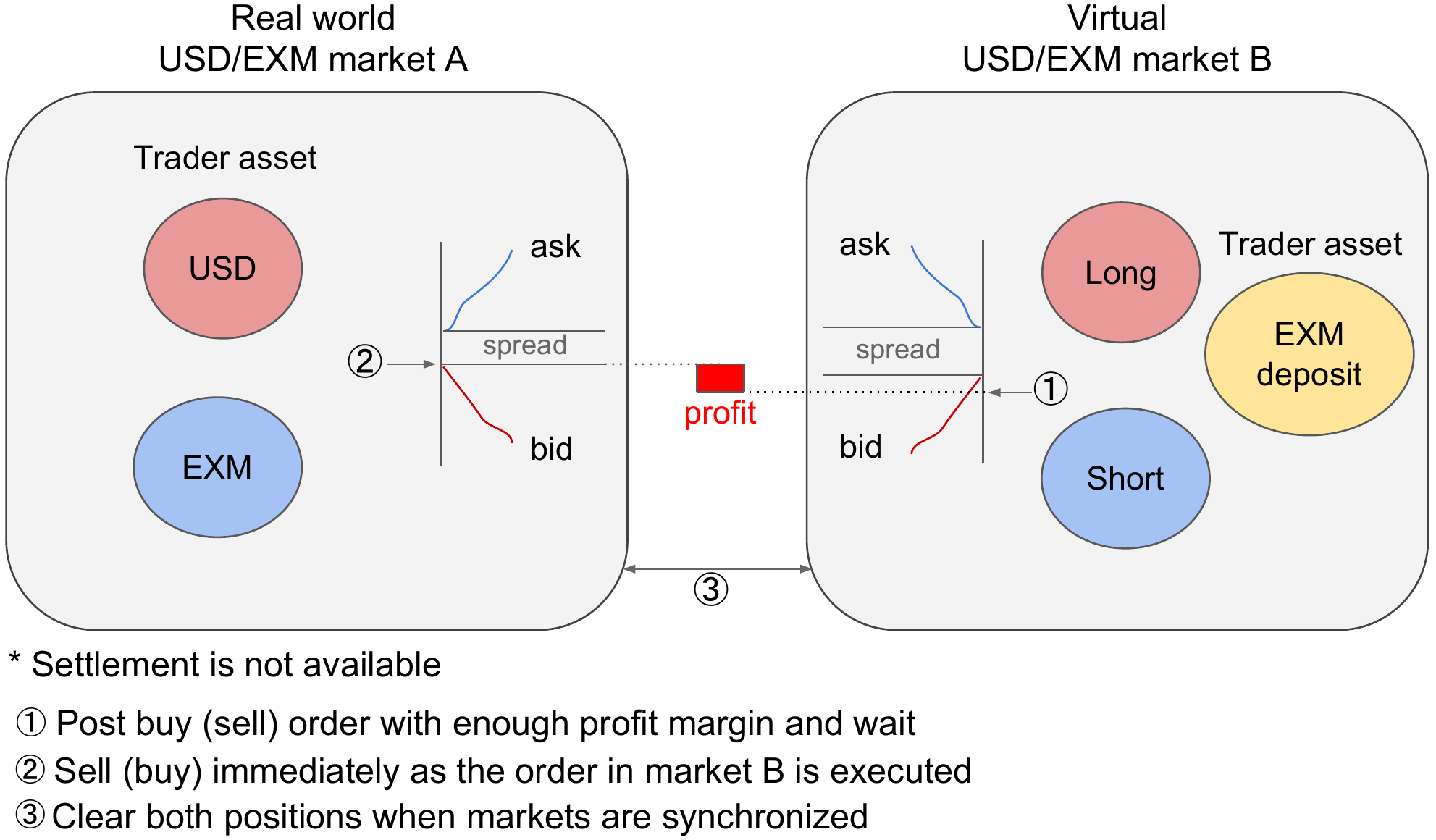}
  \caption{Statistical cross market arbitrage}
  \label{fig:arbitrage}
\end{figure}

The software performs statistical cross market arbitrage between real world markets and the virtual asset exchange. Since virtual assets are not fungible with real world assets, it may better be interpreted as pairs trading \cite{pairs}, regarding virtual and real world assets as highly correlated but separate assets. Figure~\ref{fig:arbitrage} illustrates this concept. An arbitrageur can take opposite positions on two markets -- buy on the cheap side and sell on the expensive side -- and clear the positions whenever profitable. The net effect of this trade is reducing the difference of the price between two markets. Since there are mechanisms -- swap rates in Section~\ref{sec:swap_rates} and maker rebates in Section~\ref{sec:margin_trading_dapp} -- to incentivize trades protecting the peg, the software is expected to remain profitable in the long term.

Any Exeum token holder can utilize the software to join the market and take a profit from arbitrage trading. To join the market, one needs to deposit Exeum token(s) to the system and take the market risk. The profit from the arbitrage trading should exceed the cost of taking the market risk by holding Exeum. Fortunately, it is expected that the higher the market risk, the more opportunities there are for arbitrage due to the volatility. However, if the market is too stable, there may not be enough revenue for arbitrageurs to lock their assets and wait for the volatility; in that case, the system can increase the maximum leverage ratio to improve the ROI of arbitrageurs. 

\begin{figure}
  \centering
    \includegraphics[width=\textwidth]{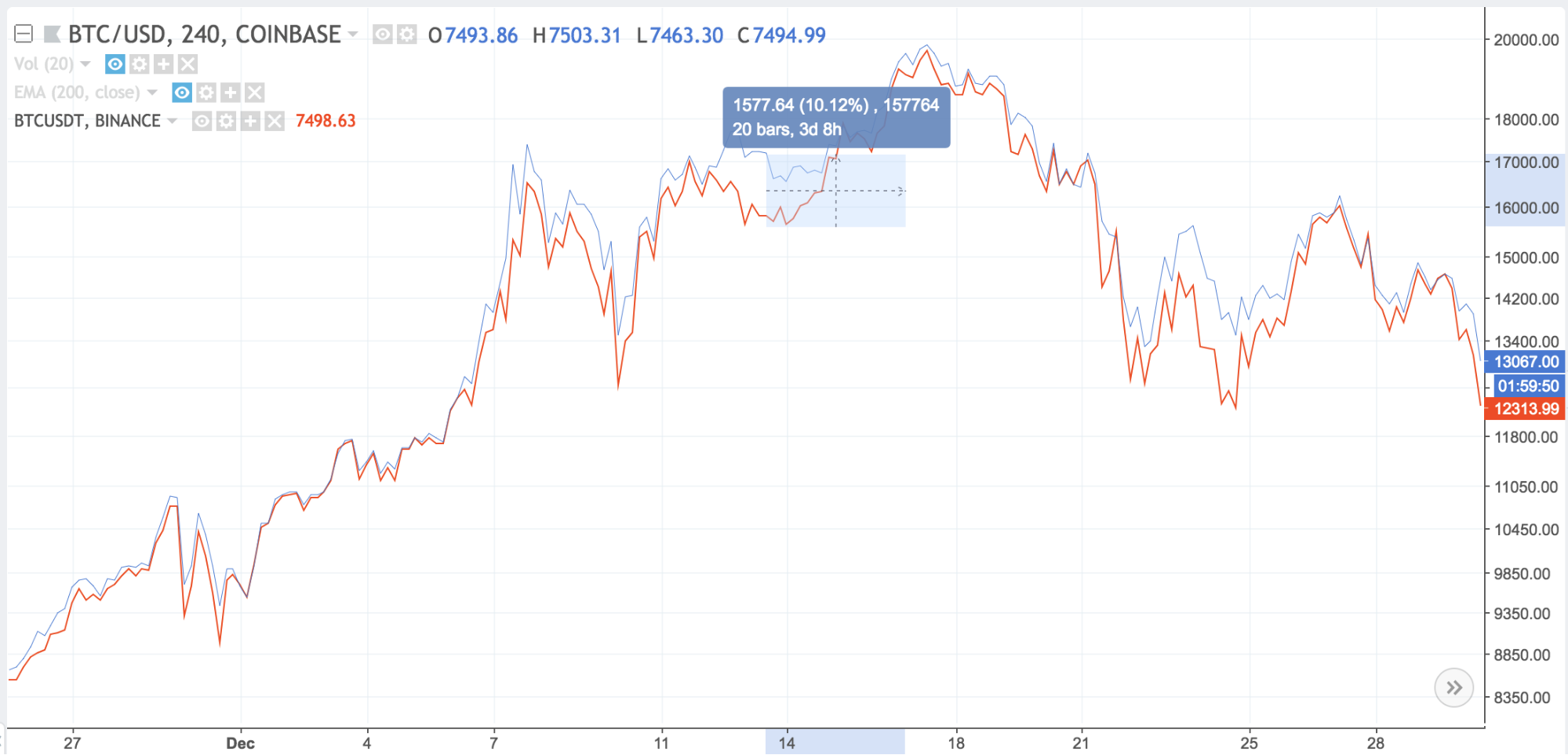}
  \caption{BTC Price comparison of Coinbase and Binance (Dec 2017)}
  \label{fig:btchistory}
\end{figure}

In practice, the price of crypto assets in USD is different by up to 10\% from exchange to exchange, as illustrated in Figure~\ref{fig:btchistory}, mostly because of the difference in the process of making a deposit in USD on those exchanges. For that reason, it is important to define arbitrage in a statistical sense and remain profitable under uncertainty. Ideally, a machine learning based offset prediction algorithm should be incorporated into the software to minimize the risk. More details will be discussed in Section~\ref{sec:strategy_arbitrage}.

\subsubsection{Obtaining Market Price Feed}
\label{sec:price_feed}
An additional role of the Arbitrage Mining Software is delivering the market price feed to the Virtual Asset Exchange DApp. The software regularly computes the index price, which is a volume weighted average of the prices in major crypto exchanges, for arbitrage trading. To deliver the index price to the DApp, the software periodically (e.g. every 10 minutes) notifies the DApp of its availability. The DApp randomly samples dozens of user addresses every minute (or more frequently if necessary) and sends price requests to them to obtain the index price feed. The software subscribes to the request and returns the computed index price along with the timestamp. The DApp verifies the submitted result by using a Schelling point scheme \cite{schelling} and rewards honest addresses.

Since each arbitrage miner deposits a certain amount of Exeum tokens, the DApp can penalize dishonest nodes by slashing the deposit. For better safety, the DApp can sample the accounts with deposits that exceed a certain threshold only. Due to the required deposit and random sampling, Sybil attacks can be prevented efficiently.

\subsection{Market Maker DApp}
\label{sec:market_maker}

Market makers place buy and sell orders around the quoted price utilizing their inventory, hoping to make a profit from the bid-offer spread. This essential activity reduces the bid-offer spread, provides liquidity on the market, and eventually benefits all market participants. In case of the virtual asset exchange, having adequate amounts of margin deposits enables market making on all the assets in the exchange. The proposed system creates a reserve of Exeum tokens at the genesis block and makes it only accessible by the Market Maker DApp, a decentralized application performing market making on the virtual asset exchange. The size of the reserve can be arbitrarily large, e.g. 2x of the total liquid amount. Aside from its normal market making activities of providing liquidity around the market price with an adaptive logic to remain profitable, the Market Maker DApp monitors the index price provided by Arbitrage Mining Software and places long and short orders around it to loosely protect the peg.

The Market Maker DApp and the reserve play an important role in the system by allowing market makers to be more aggressive taking the arbitrage and maintaining the peg. Note that one of the reasons for creating a new base token rather than using the native platform token for deposit is to create an initial reserve of a sufficient size -- otherwise a significant amount of initial capital is required to create the reserve, even more capital needs to be gathered as the system grows, and it is difficult to grow the size of the reserve as necessary without a native token reserve. Creating the reserve with a new base token permanently guarantees that there is enough liquidity for market making, since the value of the base token in the reserve increases in proportion to the amount of deposit in the system.

By design, the Market Maker DApp should remain profitable over time as market makers in real world markets do. The fee policy of the virtual asset exchange is favorable to the Market Maker DApp in that takers pay a rebate to makers. In the case that the Market Maker DApp turns out to be too profitable, Exeum token holders benefit from it since the tokens absorbed to the reserve is effectively removed from circulation.

Note that although the Market Maker DApp may do a reasonable level of market making, it cannot protect the peg by itself without accessing real world markets -- it is impossible for the Market Maker DApp to maintain the peg and protect itself from market manipulation at the same time. Attackers who have access to real world markets can manipulate real world market prices and profit from the virtual asset exchange. To counter this type of attack, the system relies on arbitrage miners or other external market makers to protect the peg and focuses on providing the right tools and incentives for them.

\subsection{Central Bank DApp}
\label{sec:central_bank}

The Central Bank DApp determines the long term interest rate of Exeum token based on the historical interest rates used in the swap process by the Virtual Asset Exchange DApp, accepts Exeum token deposit from users with a staking period, and pays the interest using the profit from exchange operation or by inflating Exeum token supply. Traders can use their deposits in the Central Bank DApp as margin on the virtual asset exchange. A properly set interest rate can make fully backed long positions of virtual assets free from EXM swap fees in the long term, thus removing the need for stability fees on pegged tokens backed by them.

To receive interest, traders need to lock their deposit for a certain period of time. Those who do not want a long term commitment can choose not to lock the deposit and give up the interest. Higher interest rates will encourage more traders and market makers to hold EXM, as well as increase the number of arbitrage miners since locked deposits can also be used for arbitrage mining. High EXM interest rates are likely to encourage USDE holders to convert their USDE to EXM, and lower interest rates will encourage the reverse, thus allowing the Central Bank DApp to control the demand of USD/EXM long positions in the system.

The long term interest rate can be adjusted taking various aspects of the system into consideration, but there is one condition it should meet -- the profit of the Currency Service DApp generated from long term interest of its EXM deposits should always be greater than the loss from the swap process, averaged in the long term. Meeting this condition guarantees that the Currency Service DApp does not lose capital in case the short term interest rate is not in favor of the DApp. 

The Central Bank DApp may also pay interest to pegged tokens such as USDE if it turns out to be beneficial for the ecosystem, but the implications can be complicated.

\subsection{Currency Service DApp}
\label{sec:currency_service}

By default, virtual asset contracts in the Virtual Asset Exchange DApp are not fungible and cannot be used by other DApps or traded in real world exchanges. The Currency Service DApp fills this gap by taking positions on the virtual asset exchange on behalf of users and issuing fungible tokens collateralized by the aforementioned positions -- it receives EXM from users, makes a deposit on the exchange, and takes a fully backed long position of the requested asset. Tokens issued by the Currency Service DApp are standard platform tokens (such as ERC20) which are general purpose. Purchasing an asset in the virtual asset exchange -- i.e. taking a long position of USD/EXM contract -- is equivalent to selling EXM in real world exchanges to obtain USD. The DApp can completely hedge itself using a long position of USD/EXM when issuing a USD-pegged token, USDE, to users. However, in doing so, it is exposed to potential loss due to the swap process (as explained in Section~\ref{sec:swap_rates}), which is compensated for by receiving long term interest on the EXM deposit (as explained in Section~\ref{sec:central_bank}).

The Currency Service DApp provides both a web user interface and a set of APIs to receive Exeum tokens and issue real world asset pegged tokens, and vice versa. The conversion is done at market rates for immediate conversion requests. Users pay a small amount of fees to the DApp to cover the exchange fee and the operational cost of the DApp. For user convenience, the DApp also supports delayed conversion of a large amount of fiat pegged tokens, by limiting the exchange rate or specifying the maximum time delay. Users can monitor the status of their conversion request via web UI or API request.

From an end user perspective, it may be desirable to enable a direct purchase of USDE using platform tokens such as Ethereum. For this purpose, the virtual asset exchange can handle one exceptional token pair -- EXM and the platform token. For example, the exchange will handle native EXM/ETH pair and let the Currency Service DApp utilize it for the direct conversion between ETH and pegged tokens. Trading this pair does not require any deposit since the exchange is done natively.

Note that the Currency Service DApp is free from liquidation risk since a fully backed long position of USD/EXM has exactly the same amount of margin as the maximum loss from the position. The maximum loss occurs when the value of USD becomes zero. In this case, the difference between the initial and final value of the position is the same as the amount of margin, and the trader loses her whole deposit when she clears the position.

\section{Stability Analysis}
\label{sec:stability}
Since fiat pegged tokens provided by Exeum are collateralized by an equivalent amount of virtual assets, guaranteeing the stability of the pegged tokens boils down to the problem of maintaining the peg of the virtual assets to real world assets.

Arbitrage miners maintain the peg of the virtual asset exchange by grabbing the arbitrage opportunities and transforming any peg breaking into their profit. The peg restores as long as there are enough arbitrage taking activities in the market. Arbitrage miners can continue their operation until their positions are skewed to the point that they run out of the base token deposit on the virtual asset exchange or funds in real world exchanges, therefore, it is important they balance their capital on different exchanges over time to maximize their profit.

If, for some reason, the market shifts drastically in one direction within a short timeframe, it is possible that existing arbitrage miners run out of capital and the peg breaks. If the peg does not restore in a reasonable time frame, it could cause significant panic selling by pegged token holders or EXM holders. If the system enters a negative feedback loop and permanently loses the peg, it should be regarded as unstable. In this section, we assume this scenario and analyze whether the system can restore the peg over time by providing enough motivation to new market participants.

Controlling swap rates is the main mechanism that the system provides for restoring the peg. If the peg breaks between the virtual asset and the real world asset, the Swap Processor in the virtual asset exchange increases the interest rate of the asset whose price is lower than the real world one, and decreases the interest rate of the asset whose price is higher. This results in the increase of the probability that market participants can profit from statistical arbitrage trading. In the following analysis, we model the difference of the prices between the virtual asset and the real world asset as a random variable whose distribution is determined by the collective prediction of future prices by market participants, and show that it is always possible to make arbitrage trades look profitable for the majority of participants by controlling the swap rate.

Without loss of generality, we focus on the USD/EXM pair on both the virtual and real world exchanges. Let the current price of USD on the real world exchange and on the virtual asset exchange be $a_0$ EXM and $b_0$ EXM, respectively. The spread $d_0$ between two exchanges is given as $(b_0-a_0)/a_0$. Let the future price (after time $t$ has passed) of USD be $a_t$ and $b_t$ EXM on the real world and the virtual asset exchange, respectively. The spread $d_t$ is calculated as $(b_t-a_t)/a_t$ in this case. The swap rate (the difference between USD and EXM interest rate) is given as $r_t$. When the swap occurs, the actual swap amount is the multiplication of the asset price $a_t$ and the swap rate $r_t$ (i.e. total $a_t r_t$ EXM). $r_t$ can be either a positive or a negative number, and a positive $r_t$ implies that short position holders of USD/EXM pays net interest to long position holders (we assume the borrowing rate and the lending rate are the same in this analysis). The arbitrage opportunity takes place if the expected profit from the statistical arbitrage trading is higher than the sum of risk-free return calculated from the risk-free interest rate of the assets used in the trading. Let the risk-free return per unit asset during the period $t$ be $R_D$ for USD, and $R_E$ for EXM.

\subsection{Case 1: Negative Spread}

\label{sec:stability_virtual_low}
\begin{table}[H]
    \centering
    \begin{tabular}{ | p{0.05\textwidth} | p{0.4\textwidth} | p{0.4\textwidth}  |}
    \hline
         & Present $(0)$ & Future $(t)$  \\ \hline
    USD Price 
            & real world exchange: $a_0$ EXM \newline virtual asset exchange: $b_0$ EXM 
            & real world exchange: $a_t$ EXM \newline virtual asset exchange: $b_t$ EXM
    \\ \hline
    Trade
            & real world exchange: \newline sell $b_0/a_0$ USD to obtain $b_0$ EXM \newline transfer $b_0$ EXM to virtual asset exchange \newline virtual asset exchange: \newline take a long pos. of 1 USD using $b_0$ EXM as margin
            & virtual asset exchange: \newline earn $a_t r_t $ EXM from the swap \newline clear the long position to earn $b_t$ EXM \newline transfer $a_t r_t + b_t $ EXM to real world exchange \newline real world exchange: \newline sell earned EXM to obtain $r_t + b_t/a_t$ USD
    \\ \hline
    Asset flow & $-b_0/a_0$ USD & $r_t + b_t/a_t$ USD \\
    \hline
    \end{tabular}
    \caption{Arbitrage trading strategy in the case USD price in the virtual asset exchange is lower than the one in the real world exchange. Asset flow is measured in USD.}
    \label{tab:case1}
\end{table}

The arbitrage trading strategy in the case USD price in the virtual asset exchange is lower than the one in the real world exchange (that is, $d_0 < 0$) is described in Table~\ref{tab:case1}. First, the trader sells $b_0/a_0$ USD to obtain $b_0$ EXM and deposit it on the virtual asset exchange as a margin. Using the margin, she takes a fully backed long position of 1 USD (without liquidation risk). Total asset flow from the activities above is $-b_0/a_0$ USD.

Next, after time $t$ has passed and the swap occurs, she clears the long position. $a_t r_t$ EXM is earned from the swap, and $b_t$ EXM is earned from the cleared position. This EXM is moved to the real world exchange and sold to obtain $(r_t+b_t/a_t)$ USD. From the discussion above, the arbitrage opportunity takes place if:
\begin{equation}
\label{eq:case1}
r_t + b_t/a_t > (b_0/a_0)(1+R_D).
\end{equation}

We can rewrite the condition above using the spread variables $d_0$ and $d_t$ as below:
\begin{equation}
\label{eq:case1a}
r_t > d_0 - d_t + R_D(1+d_0),
\end{equation}
which is independent of the market's direction. Thus, this is a market neutral strategy. Note that $-1 \leq d_0 < 0$ and $-1 \leq d_t < \infty$. $d_t = 0$ implies that the peg has been restored, and $d_t > 0$ implies that the market condition has reversed -- USD price on the virtual asset exchange is higher. Higher $r_t$ is required for the arbitrage opportunity as $d_t$ approaches $-1$, but it is always possible to find an $r_t$ value which satisfies Eq.~\eqref{eq:case1a}.

On the other hand, Eq.~\eqref{eq:case1a} can be rearranged with respect to $d_t$:
\begin{equation}
\label{eq:case1b}
d_t > -r_t + d_0 + R_D(1+d_0).
\end{equation}

\begin{figure}[H]
\begin{center}
\subfloat[]{\includegraphics[width=0.45\textwidth]{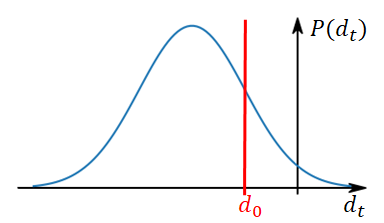}\label{fig:case1a}}
\subfloat[]{\includegraphics[width=0.45\textwidth]{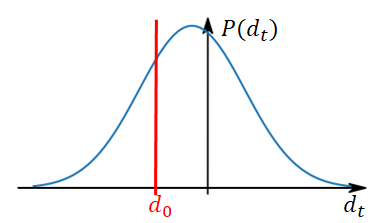}\label{fig:case1b}}
\caption{Probability distribution of the predicted spread $d_t$ at time $t$. (a) Market participants predict that $d_t$ will decrease over time. (b) Market participants predict that $d_t$ will increase over time.}
\label{fig:case1}
\end{center}
\end{figure}

We can model $d_t$ as a random variable which implies the future prediction of the spread at time $t$ by market participants. Let $P(d_t)$ be the probability distribution of $d_t$. Figure~\ref{fig:case1} shows two different examples of $P(d_t)$. Figure~\ref{fig:case1} (a) corresponds to the case market participants predict that $d_t$ will decrease over time (that is, $d_t$ approaches -1). Figure~\ref{fig:case1} (b) corresponds to the case it is predicted that $d_t$ will increase over time (that is, $d_t$ approaches 0 and beyond). 

\begin{figure}[H]
\begin{center}
\subfloat[]{\includegraphics[width=0.45\textwidth]{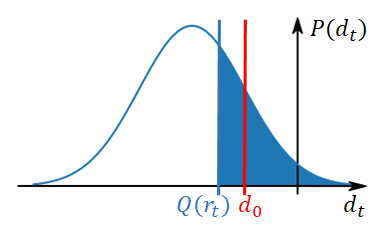}\label{fig:ratioa}}
\subfloat[]{\includegraphics[width=0.45\textwidth]{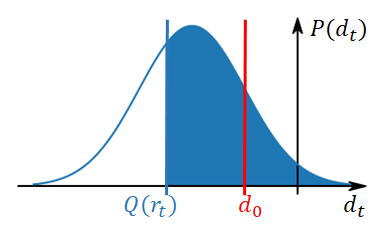}\label{fig:ratiob}}
\caption{The ratio of participants who think that the arbitrage opportunity took place (blue area), with respect to the change of the swap rate $r_t$. (a) $r_t$ is low. (b) $r_t$ is high.}
\label{fig:r_t}
\end{center}
\end{figure}

The ratio of participants who think that the arbitrage opportunity took place can be obtained by integrating $P(d_t)$ over the range that Eq.~\eqref{eq:case1b} holds true. Let the right hand side of Eq.~\eqref{eq:case1b} be $Q(r_t)=-r_t + d_0 + R_D(1+d_0)$, which is a function of $r_t$. The ratio of the participants can be obtained by $V(r_t)=\int_{Q(r_t)}^{\infty} P(x) dx$, which corresponds to the blue area in Figure~\ref{fig:r_t}. A higher value of $r_t$ lets more participants find an arbitrage opportunity. If more than half of the participants find an arbitrage opportunity, the price difference between two exchanges will decrease due to arbitrage trades. It can create a positive feedback loop since it can alter the distribution $P(d_t)$ from Figure~\ref{fig:case1} (a) to Figure~\ref{fig:case1} (b), eventually restoring the peg. Since $r_t$ is fully controllable by the system, it can be gradually increased until enough amount of arbitrage trade occurs.

\subsection{Case 2: Positive Spread}
\label{sec:stability_virtual_high}

\begin{table}[H]
    \centering
    \begin{tabular}{ | p{0.05\textwidth} | p{0.42\textwidth} | p{0.42\textwidth}  |}
    \hline
         & Present $(0)$ & Future $(t)$  \\ \hline
    USD Price 
            & real world exchange: $a_0$ EXM \newline virtual asset exchange: $b_0$ EXM 
            & real world exchange: $a_t$ EXM \newline virtual asset exchange: $b_t$ EXM
    \\ \hline
    Trade
            & real world exchange: \newline sell $b_0/(1+R_E)$ EXM to obtain $b_0/(a_0(1+R_E))$ USD \newline virtual asset exchange: \newline take a short pos. of 1 USD (deposit enough margin to avoid liquidation)
            & real world exchange: \newline sell $b_0/(a_0(1+R_E))$ USD to obtain $a_t b_0/(a_0(1+R_E))$ EXM \newline virtual asset exchange: \newline earn $a_t |r_t|$ EXM from the swap \newline clear the short position to earn $b_0 - b_t$ EXM
    \\ \hline
    Asset flow & $-b_0/(1+R_E)$ EXM & $ a_t(b_0/(a_0(1+R_E))+|r_t|) + b_0 - b_t$ EXM \\
    \hline    
    
    \end{tabular}
    \caption{Arbitrage trading strategy in the case USD price in the virtual asset exchange is higher than the one in the real world exchange. Asset flow is measured in EXM.}
    \label{tab:case2}
\end{table}

The arbitrage trading strategy in the case USD price in the virtual asset exchange is higher than the one in the real world exchange is described in Table~\ref{tab:case2}. It is assumed that the trader holds a certain amount of EXM as a deposit in Central Bank DApp, which is enough for taking necessary positions in the virtual asset exchange for arbitrage trading. Note that this EXM deposit is not included in the risk-free return calculation since Central Bank DApp pays interest on it. 

First, the trader sells $b_0/(1+R_E)$ EXM to obtain $b_0/(a_0(1+R_E))$ USD. At the same time, she takes a short position of 1 USD. Total EXM asset flow from the activities above is $-b_0/(1+R_E)$ EXM. Next, after time $t$ has passed and the swap occurs, she clears the short position. $a_t |r_t|$ EXM is earned from the swap, and $b_0 - b_t$ EXM is earned from the cleared position. At the same time, she sells $b_0/(a_0(1+R_E))$ USD in the real world exchange to obtain $a_t b_0/(a_0(1+R_E))$ EXM. The total asset flow is $a_t(b_0/(a_0(1+R_E)) + |r_t|) + b_0 - b_t$ EXM. From above, the arbitrage opportunity takes place if:
\begin{equation}
a_t\bigg(\frac{b_0}{a_0(1+R_E)}+|r_t|\bigg) + b_0 - b_t> \frac{b_0}{1+R_E}(1+R_E),
\end{equation}
which can be rearranged using the spread variables $d_0$ and $d_t$ as below:
\begin{equation}
|r_t| > d_t - \frac{d_0}{1+R_E} + \frac{R_E}{1+R_E}
\end{equation}
which is independent of the market's direction. Similar to Section~\ref{sec:stability_virtual_low}, a higher value of $|r_t|$ lets more participants find an arbitrage opportunity, and the system can control $r_t$ to increase the amount of arbitrage trade and restore the peg.

\section{Comparison with Other Approaches}
\label{sec:comparison}

In this section, we examine several existing approaches to developing price stable cryptocurrencies and compare them with our proposed approach.

\subsection{Collateral Backed -- Centralized IOU Based}

In this approach, one or more centralized companies hold users’ fiat deposit in a bank account and issue an equivalent amount of digital currency backed by the deposit, and vice versa. Tether dollars, issued by a single company named Tether~\cite{tether}, has been the most successful approach to stablecoins thus far, having been adopted by major crypto exchanges for years without major issues despite serious counterparty and regulation risks. The price of tether dollars has remained relatively stable, although there have been multiple hacking incidents and regulation issues \cite{tetherhack}.

\begin{figure}
  \centering
    \includegraphics[width=0.8\textwidth,height=0.45\textheight]{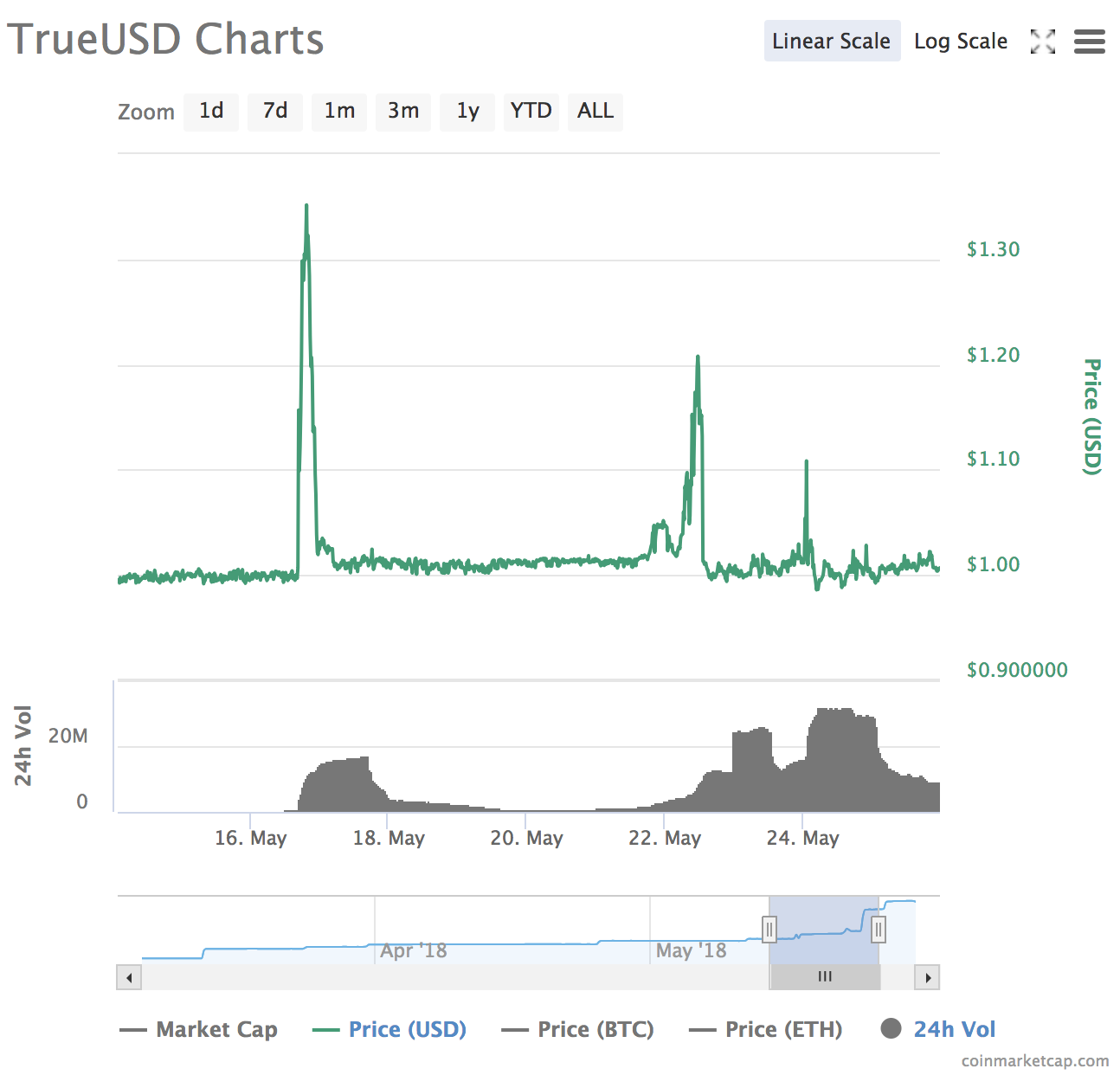}
  \caption{Price history of TrueUSD}
  \label{fig:trueusd}
\end{figure}

TrueUSD~\cite{trueusd}, a new alternative to Tether which significantly reduces the counterparty risk by utilizing multiple trust companies and escrow accounts (although regulation risk still persists), suffers from price instability (as in Figure~\ref{fig:trueusd}). 

The relative difference in observed stability between Tether and TrueUSD implies that proper collateralization does not always guarantee price stability because market prices are determined by supply and demand. The takeaway is that for any approach to developing a stablecoin to succeed, a systematic market making activity must be an essential component to the system. Without it, the price can be manipulated at critical moments and businesses relying on stability of the currency can be exploited. Note that Tether is also backed by one of the largest crypto exchanges, Bitfinex, and there is likely to be active, professional market making activity to maintain the stability of Tether dollars, which is crucial to its success.

In this respect, the limited supply issue due to potential regulatory risks, as pointed out in Section~\ref{sec:intro}, becomes critical. Market makers would need access to an additional stablecoin supply depending on the market situation. If the supply is limited due to regulatory changes, there is practically no way to keep the price stable. Thus, a decentralized issuance mechanism is crucial for the long term price stability of a stablecoin, and IOU-backed models should be regarded as a temporary measure despite its current success.

\subsection{Collateral Backed -- Crypto Asset Based}

\begin{figure}
  \centering
    \includegraphics[width=0.8\textwidth,height=0.45\textheight]{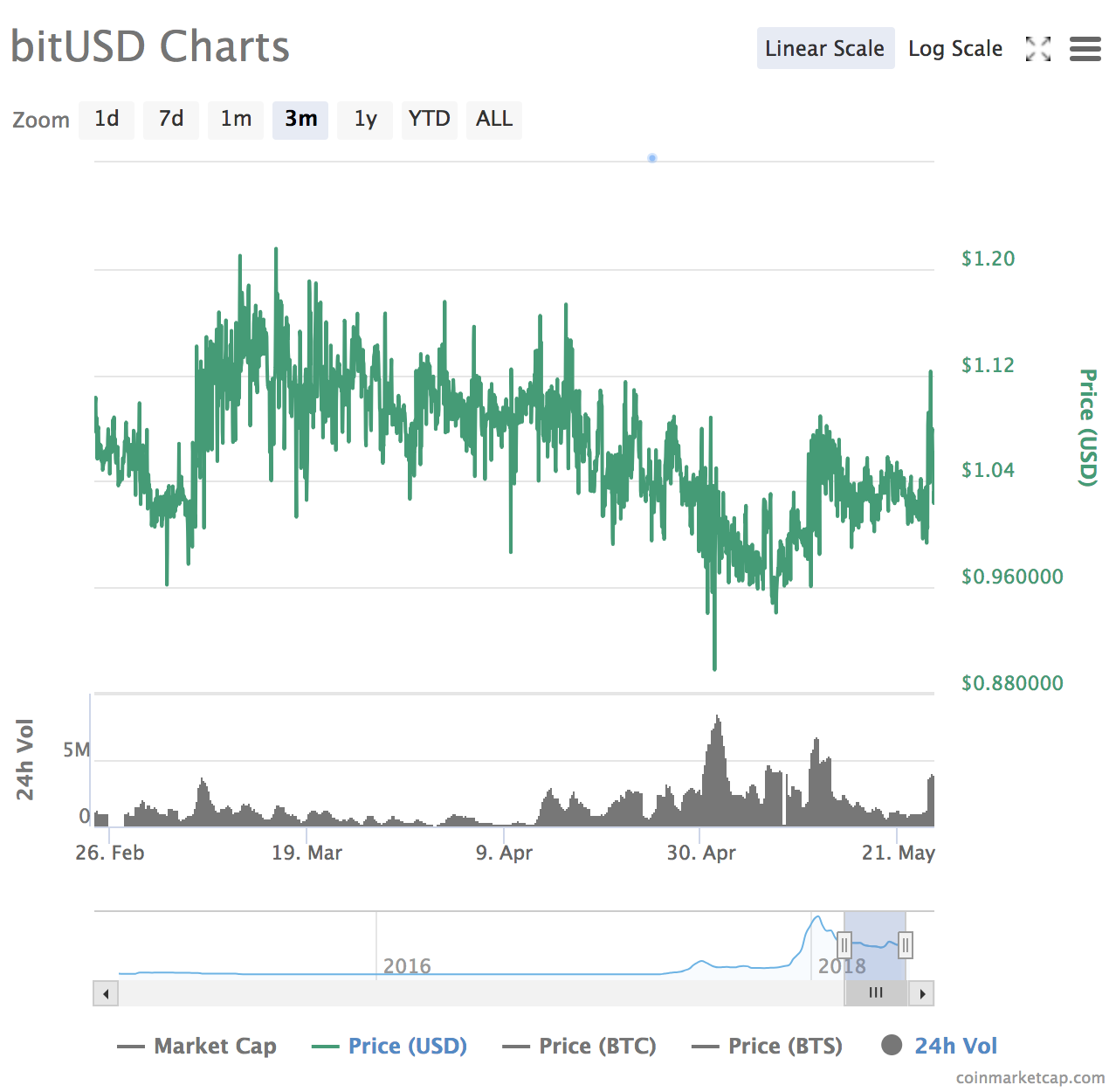}
  \caption{Price history of bitUSD}
  \label{fig:bitusd}
\end{figure}

\begin{figure}
  \centering
    \includegraphics[width=0.8\textwidth,height=0.45\textheight]{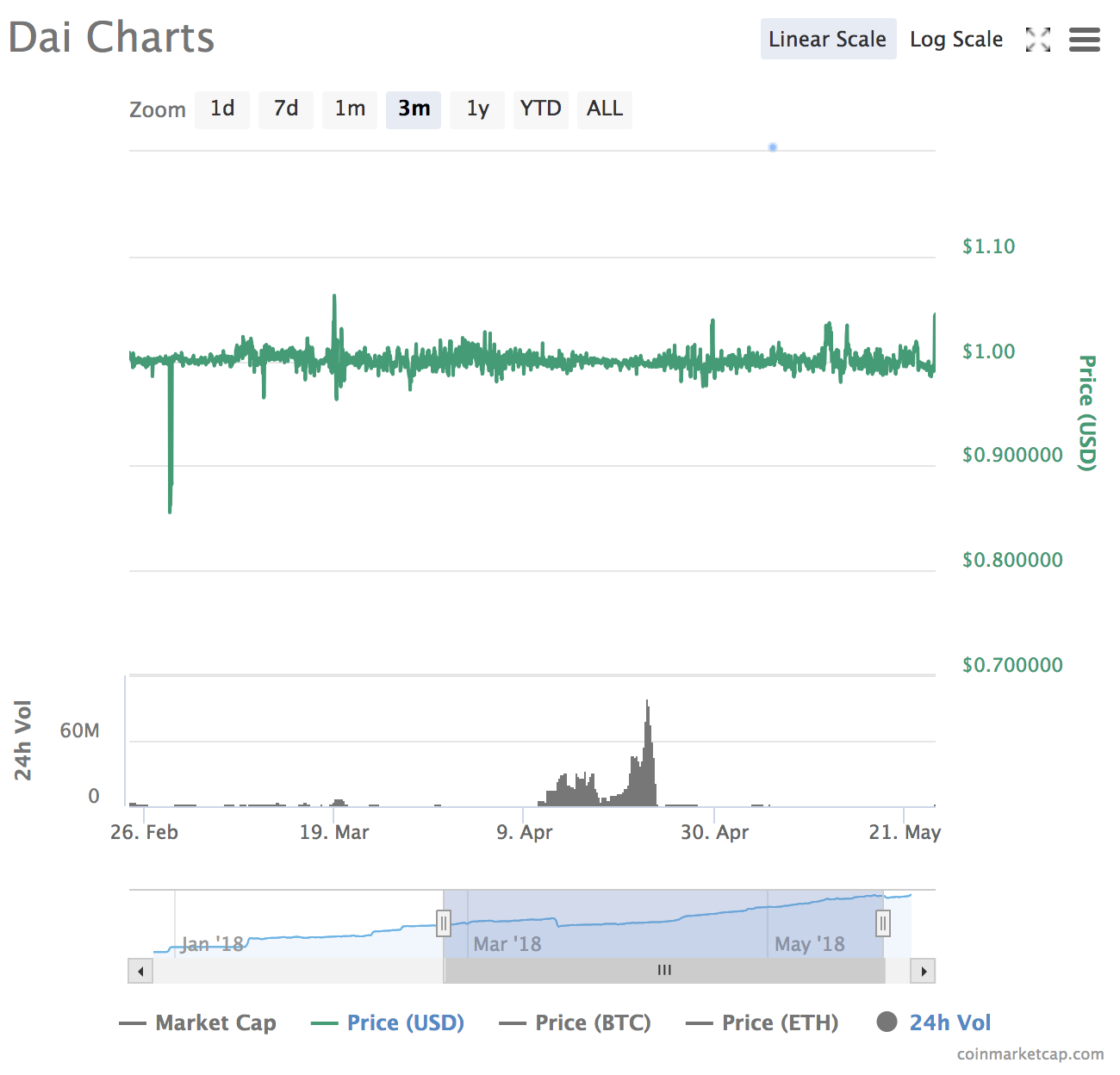}
  \caption{Price history of DAI (MakerDAO)}
  \label{fig:dai}
\end{figure}

A more decentralized approach to collateral backed stablecoins is backing them by other decentralized crypto assets. Due to the volatility of crypto assets, existing stablecoins using this approach (e.g. BitShares \cite{bitshares} and MakerDAO \cite{dai}) are usually over-collateralized (2x) to overcome movements in price. A major issue shared by this type of approach is the lack of supply and corresponding price fluctuations. As can be seen in Figure~\ref{fig:bitusd} and \ref{fig:dai}, it is not uncommon to see a price surge of 5-20\% when there is a strong need for stablecoins, such as when the crypto market crashes. The surge occurs because the token supply mechanism is isolated from demand (i.e. token is not automatically issued based on demand -- only suppliers can decide to issue stablecoins and sell them on the market when they want to). To overcome this problem, there needs to be a strong incentive to market makers so that they are willing to monitor the market and issue coins instantly when there is a need. In earlier systems like BitShares, this was accomplished by adding 10-20\% premium to the stablecoin price (by itself breaking the price stability). Improved systems like MakerDAO tries to adjust the supply by dynamic fee mechanisms including a stability fee, but the response cannot be immediate as we can see from the price graph.

Another problem of this approach is a `black-swan’ event, which happens when the price of collateral falls below the settlement threshold. BitShares suffers from frequent `black-swan' events even though its stablecoin is backed by a derivative contract. There is no derivatives market (although there is a decentralized exchange) in BitShares and the only counterparty of the contract is the system itself. If we replicate the BitShares approach in Exeum, it would be equivalent to only allowing the Currency Service DApp to take long positions. As a result, if any of the short positions gets liquidated in BitShares, it would be comparable to the Currency Service DApp being forced to clear some of its long positions. As issued stablecoins are fungible, the only way to remain fair without a derivatives market is to clear all the long positions (hence all the bitUSD) and declare a 'black-swan' event, which exposes stablecoin holders to market risks immediately.

In case of MakerDAO, this 'black-swan' event is handled by opening up an auction immediately and selling the liquidated position to participants. In practice, this means that only the bots can monitor the system and participate in the auction, leading to potential centralization and market manipulation.

Exeum resolves the limited supply issue above by allowing the issuance of stablecoins upon user request -- in this way, the demand controls the supply directly. A sudden high demand of stablecoins pushes the value of Exeum token up, but does not affect the price of the pegged tokens -- e.g. USDE in this case. The 'black-swan' event is also handled gracefully since the liquidation of a short position does not need to be handled explicitly by the system and is absorbed by other market participants.

\subsection{Models Based on Seigniorage Shares}

The latest approach to stablecoins are Seigniorage share \cite{seign} based models which algorithmically expands and contracts the supply of the stablecoin similar to the function that a central bank performs for fiat currencies. Shareholders of the system are rewarded with newly issued stablecoins when the supply expands, and the system contracts the supply by issuing new shares or so called ‘bonds’, which are paid back when there is an expansion. 

Basis~\cite{basis}, the first to propose such an approach, relies entirely on the quantity theory of money~\cite{moe} for price stabilization. Although Basis is not on the market yet, based on observations of other stablecoins, the market price of Basis will likely have short term fluctuations unless major market makers play a role and stabilize the price. In other words, Basis shareholders need to actively perform market making activities to maintain its price. The caveat is that if the market making activity is too strong, it affects the measurement mechanism of Basis -- the algorithm cannot expand or contract the supply since the price does not fluctuate. 

Another problem is potential centralization of the system -- since there is no built-in incentive mechanism or tools provided for market making, it is not profitable for minor shareholders to join market making activity. The likely result is that market making is dominated by major shareholders of Basis given the long term incentives. Once centralized, market makers (i.e. major shareholders) can induce artificial expansions and contractions of supply at unexpected times, forcing the system to distribute expanded amounts of Basis-USD to shareholders and let them use it to buy bonds when an artificial contraction happens. Since Basis shares are not involved in the process, shareholders take no risk of losing their shares in the process and profit as long as they can purchase bonds at a cheaper price, opening exploit opportunities in the long term.

New projects like Carbon~\cite{carbon} create a reserve by setting aside a portion of the supply expansion. The system uses the reserve to be involved in the market directly in order to remedy the short term fluctuation issue, which, as analyzed above, affects the ability of the system to measure supply contraction and expansion. A naive implementation of the market maker would cause the reserve to deplete before any supply expansion happens. A clever algorithm may be able to determine when to stop the involvement and start supply expansion rather than depleting the reserve, but it is not easy to predict all the potential attack vectors to profit from the reserve operation.

One common problem all the Seigniorage share based models would confront over time is that there is no rescue when the system starts a chronic deflation. There is no economy which lasts forever -- many countries and their fiat currencies have disappeared from history over time. Decentralized economies on which stablecoins operate will not be an exception. Assuming deflation eventually occurs, the share or bond of Seigniorage share based systems would have zero value and not be able to contract stablecoin supply by buying them back from the market. In other words, they suffer from a circular reference of value backing (i.e. their shares or bonds have value as long as the peg works, and the peg works only if the shares or bonds have value). 

Exeum takes these issues into account and does not rely on a fixed algorithm for supply contraction and expansion -- market demand directly controls the supply. Since supply contraction is also done on demand, Exeum does not suffer from the problems of chronic deflation, although the value of the base token will decrease along with softening demand.

\section{Applications}

Enabling price stable assets in the decentralized economy opens up endless opportunities -- payment networks, financial derivative markets, a decentralized store of value, etc -- most of which have been envisioned and elaborated upon by other stablecoin projects. In this section, we focus on introducing a number of applications of the proposed service that are not as obvious.

\subsection{Virtual Crypto Assets Exchange}

At the moment, the major use case for stablecoins is to provide USD alternatives on centralized crypto exchanges. Using pegged assets as quote assets, the Virtual Asset Exchange DApp provided by the Exeum project can be expanded to support major crypto pairs such as BTC/USDE, ETH/USDE, ETH/BTCE, and so on. Users can deposit fiat pegged assets to be free from crypto volatility risks and actively engage in trading. Since these additional pairs do not affect the stability of the pegged token, the exchange DApp can allow high degree of leverage for them to make it attractive to traders. The value of Exeum tokens will be supported by increased trading activity on those pairs, and arbitrage miners will benefit from increased market opportunities.

\subsection{Hedging Staking-based DApp Token Models}

Recent proposals for utility token models require participants to stake tokens to earn various rights from the service. The rise of such models is inevitable given that there are only two ways of increasing the value of a token from the quantity theory of money point of view -- controlling the supply, or reducing the velocity of the token \cite{moe}. Given that most of the tokens have a fixed supply, the only way to increase the value of the tokens is to reduce the velocity, which can be done by requiring participants to stake them in return for the benefits provided by the network. The most classic example is Steem \cite{steem} -- users need to stake tokens -- i.e. converting STEEM to STEEM POWER -- to gain more influence on the system. There are a number of projects which require service providers to stake DApp tokens to earn the right to perform work for the network \cite{worktoken} -- for example, Keep \cite{keep}, a secure enclave service on the Ethereum network, requires service providers to stake KEEP tokens in order to receive ETH payments from users. 

While the aforementioned improved utility token models may help increase the value of the token in proportion to the usage of the network, participants are burdened with direct exposure to the volatility of DApp tokens. In turn, this effectively restricts the group of participants to strong believers of the token’s future value, which limits broader adoption of the service. The Exeum service can provide a solution to this issue by allowing users to hedge their stakes using the virtual asset exchange -- effectively providing means to separate the role of investors and workers.

\subsection{DApp Platform with Price Stable Gas Fee}

In an ideal world, users of DApp platforms should not be exposed to any market risk -- which has always been the case for centralized services. While users can hedge themselves using the Exeum service and keep using platform tokens, there is still a remaining market risk of the gas fee fluctuating with the price of platform tokens. If the platform allows DApps to pay for gas fees directly on behalf of their users, the issue of fluctuating gas prices becomes clearer and will make long term planning of DApp based businesses difficult.

To make the experience of using DApps closer to centralized alternatives, it would be beneficial for users if the Exeum service is tightly integrated with the platform and the gas fee is paid in price stable tokens directly. DApp platforms need to be conscious of resource efficiency sooner or later to remain competitive. Operator nodes would also benefit from price stability if they start operating with a thinner margin, providing more computational resources at cheaper prices. To realize these advantages, Exeum tokens could be merged with platform tokens to maintain the value of the platform tokens -- otherwise there would be no value remaining for platform tokens. In the case of a merger, platform tokens still maintain their value since any price stable tokens should be collateralized by positions in the virtual asset exchange, which are eventually backed by platform token deposits.

\subsection{Cross-Chain Asset Transfer}

Since Exeum service can provide pegged tokens for any assets traded in real world markets, it is possible to create a token which is pegged to an asset in other blockchains. For example, a pegged token named BTCE, whose price is pegged to Bitcoin, can be created on Exeum-enabled DApp platforms. For many applications, this can be a superior alternative to other cross-chain asset transfer methods because of its decentralization, simplicity, and low processing delay. All the major legacy crypto assets which do not have a proper built-in support for cross-chain interaction can enter major DApp platforms in this way.

Looking forward, a mature cross-chain asset transfer mechanism between DApp platforms will likely be developed and will actually benefit the Exeum service since its pegged tokens can be transferred to other DApp platforms without the groundwork of opening up a separate market per platform and securing enough liquidity.

\section{Discussions}
\label{sec:discussion}
In this section, we discuss several practical aspects of the proposed system, its future direction, and its potential implications on the decentralized and real world economies.

\subsection{The Impact of Allowing Leverage}

For asset pairs used by the Currency Service DApp (e.g. USD/EXM), it is critical to maintain the peg between the virtual asset exchange and real world exchanges. Doubling the leverage amount will double the profit of arbitrage trading, but will impact the breakdown threshold. For example, let’s assume the price of USD/EXM pair is 100 (i.e. 100 EXM per USD). With 1x leverage, arbitrage miners taking the short position will get margin called if the price of USD/EXM reaches 200 (i.e. 200 EXM per USD). That is, the miner loses her 100 EXM deposit if she sold 1 USD originally to receive 100 EXM and buys it back after it reaches 200 EXM per USD, since the difference of two positions is 100 EXM. Miners can sustain a sudden surge of USD price by 100\% in this case.

With 2x leverage, she can take a short position of 1 USD with only 50 EXM. It gets margin called when the price hits 150 EXM per USD since her deposit, 50 EXM, cannot sustain more than 150-100=50 EXM difference. Miners can sustain a sudden surge of USD price by 50\% in this case. A sudden increase / decrease of EXM price that occurs too quickly for arbitrage miners' positions to be cleared may cause mass liquidation of their positions and a broken peg. For this reason, it is important to restrict maximum leverage considering the overall volatility of the system.

\subsection{Machine Learning for Arbitrage Mining}
\label{sec:strategy_arbitrage}
Since the index price on the virtual asset exchange is computed as the volume weighted average of prices on real world exchanges (which could be significantly different from each other -- refer to Figure~\ref{fig:btchistory}), a difference exists between the index price and the price on individual real world exchanges. The mining software needs to model the characteristics of this difference and optimize its behavior in a statistical sense.

While Exeum will provide completely open source software for arbitrage mining for transparency, the model and algorithm included in the open source software needs to remain conservative to reduce the possibility of exploits, which impacts profitability. To address this, Exeum is contemplating a partially closed source implementation utilizing machine learning techniques. A number of sophisticated deep learning based models can be trained with different strategies and provided as plugins to improve profitability. To restrict unexpected behaviors of trained models, the software can provide open source and configurable safety and insurance layers to enforce certain conditions on executed trades.

\subsection{Implications on Traditional Economy}

The Exeum project adds a new liquid asset class -- Exeum token, EXM -- to the real world economy, increasing the total wealth and liquidity by the total value of Exeum token. This added value can be converted to any other pegged asset, which effectively enables free conversion between currencies. In case the decentralized economy supported by Exeum tokens becomes significant, it may affect the currency policy of local governments and cause issues.

On the other hand, the amount of locked capital for issuing a pegged token is always greater than the notional value of the pegged token. For example, to issue one USDE, there needs to be one USD amount of long and the same amount of short positions taken on the virtual asset exchange, whose total margin deposit always exceeds one USD. Although issuing USDE effectively increases the amount of liquid USD, it does not increase total liquidity in the real world economy, unlike banknotes issued by conventional banks with a low cash reserve ratio. In this regard, Exeum system can be viewed as an automatically and strictly regulated implementation of free banking model \cite{freebanking}, where the privilege to supply currency is fully democratized and controlled by market demand.

\subsection{Pegging to A Basket of Virtual Assets}

The Currency Service DApp can be expanded to issue tokens backed by a basket of virtual assets to create a consumer price index.  As the decentralized economy grows, creating a token whose value is derived from a basket of DApp tokens for such an index could be a viable alternative to relying on fiat money and its centralized monetary policy in the future.

\section{Conclusion}
In this paper, we have introduced Exeum, a novel decentralized system to issue tokens pegged to real world assets, including fiat currencies. Pegged tokens are collateralized by the virtual assets traded in the decentralized virtual asset exchange provided by the system, effectively transforming the price stability problem into maintaining the peg between the virtual asset exchange and real world exchanges. The system implements a number of mechanisms to incentivize market makers and maintain the peg -- a rebate for maker orders, the swap rate adjusted based on the demand of the asset in the exchange, and loose protection of the peg by the Market Maker DApp and the initial reserve. Exeum tries to democratize market making activity by recruiting arbitrage miners, using the market making software provided by Exeum.

Users can submit Exeum tokens and request the issuance of fiat pegged tokens by the Exeum service, which are immediately collateralized by fully backed long positions taken by the system in the virtual asset exchange. The system avoids stability fees on pegged token holders by introducing a central bank equivalent and adjusting the long term interest rate on EXM deposits. Exeum is so far the only decentralized system which enables an immediate 1:1 value conversion between the base token and pegged tokens, which is designed to solve the mismatch between supply and demand problem from which both collateral based and Seigniorage share based approaches suffer.

By being tightly integrated into DApp platforms, it maximizes the possibility of being used by other DApps and forming the basis of a fully decentralized economy, with a number of interesting applications such as hedging staking-based DApp token models, cross-chain transfer of legacy crypto assets, and price stable gas fees. The Currency Service DApp can be expanded to open up the possibility to evolve the system beyond the era of fiat money, by pegging to an index of DApp tokens.

A completely decentralized stablecoin without regulation and without the constraints of a limited supply turns financial stability into a common resource -- which humanity has never experienced in its recorded history. Exeum aims to achieve this goal with optimal capital efficiency by capturing the core value of a stablecoin in a form of financial derivative, and opening a decentralized derivatives exchange to enable the efficient trading of that captured value. Efficiency has been the key ingredient in many noteworthy inventions; we believe that Exeum will deliver the gains in efficiency necessary to advance the vision of a decentralized economy.

\subsubsection*{Acknowledgements}

We would like to express our deep gratitude to Sherman Li, Dinh Duong Mai, and Dr. Yongsun Choi for detailed reviews and suggestions that greatly improved the manuscript, Youngmoo Kwon for suggesting the project name and paper title, and Jihye Ahn for improving the figures. We are grateful for helpful discussions and insightful feedback from Prof. Dongsu Han and Dr. Tae-Rog Oh, even though they may not necessarily agree with all of the interpretations and conclusions presented in this paper. Last, but certainly not least, we must extend our appreciation to the members of the Exeum team, who have been contributing their passion and intellect to the project on a voluntary basis.

\newpage
{\small
\bibliographystyle{unsrtnat}
\bibliography{exeum}
}
\end{document}